\documentclass[fleqn,usenatbib]{mnras}

\usepackage[T1]{fontenc}

\DeclareRobustCommand{\VAN}[3]{#2}
\let\VANthebibliography\thebibliography
\def\thebibliography{\DeclareRobustCommand{\VAN}[3]{##3}\VANthebibliography}

\usepackage{graphicx}	
\usepackage{amsmath}	
\usepackage{amssymb}	
\usepackage{natbib}
\usepackage{enumitem}
\usepackage[flushleft,referable]{threeparttablex}

\usepackage{newtxtext,newtxmath}

\newcommand{\LyA}{Ly$\alpha$}
\newcommand{\SSFR}{$\Sigma_{\rm SFR}$}
\newcommand{\sSSFR}{$\Sigma_{\rm sSFR}$}

\newcommand{\Vlis}{$\Delta v_{\rm{LIS}}$}

\newcommand{\Vfrc}{$v_{\rm{80}}$}
\newcommand{\Vfrcr}{$v_{\rm{80, red}}$}
\newcommand{\Vfrcb}{$v_{\rm{80, blue}}$}

\newcommand{\Vmax}{$v_{\rm{max}}$}
\newcommand{\Vmaxr}{$v_{\rm{max, red}}$}
\newcommand{\Vmaxb}{$v_{\rm{max, blue}}$}

\newcommand{\Vcenr}{$v_{\rm{cen, red}}$}

\newcommand{\RE}{$R_{\rm{E}}$}
\newcommand{\HA}{$\rm{H}\alpha$}
\newcommand{\HB}{$\rm{H}\beta$}
\newcommand{\Mste}{$M_{\star}$}
\newcommand{\Mdyn}{$M_{\rm{dyn}}$}
\newcommand{\Mgas}{$M_{\rm{gas}}$}
\newcommand{\Mbar}{$M_{\rm{bar}}$}

\DeclareUnicodeCharacter{2212}{-}

\title[MOSDEF-LRIS: Inflows]{The MOSDEF-LRIS Survey: Detection of Inflowing Gas Towards Three Star-forming Galaxies at $z \sim$ 2}

\author[A. Weldon et al.]{
Andrew Weldon,$^{1}$\thanks{E-mail: aweld004@ucr.edu}
Naveen A. Reddy,$^{1}$
Michael W. Topping,$^{2}$
Alice E. Shapley,$^{3}$
Xinnan Du,$^{4}$
\newauthor
Sedona H. Price$^{5}$,
Ryan L. Sanders,$^{6, 7}$
Alison L. Coil,$^{8}$
Bahram Mobasher,$^{1}$
Mariska Kriek,$^{9}$
\newauthor
Brian Siana,$^{1}$ and
Saeed Rezaee$^{1}$
\\
$^{1}$Department of Physics and Astronomy, University of California, Riverside, 900 University Avenue, Riverside, CA 92521, USA\\
$^{2}$Steward Observatory, University of Arizona, 933 N Cherry Ave, Tucson, AZ 85721, USA\\
$^{3}$Physics \& Astronomy Department, University of California, Los Angeles, 430 Portola Plaza, Los Angeles, CA 90095, USA\\
$^{4}$Kavli Institute for Particle Astrophysics \& Cosmology, P. O. Box 2450, Stanford University, Stanford, CA 94305, USA\\
$^{5}$Department of Physics \& Astronomy and PITT PACC, University of
Pittsburgh, Pittsburgh, PA 15260, USA\\
$^{6}$Department of Physics and Astronomy, University of California, Davis, One Shields Ave, Davis, CA 95616, USA\\
$^{7}$Hubble Fellow\\
$^{8}$Center for Astrophysics and Space Sciences, Department of Physics, University of California, San Diego, 9500 Gilman Dr., La Jolla, CA 92093-0424, USA\\
$^{9}$Leiden Observatory, Leiden University, PO Box 9513, NL-2300 RA Leiden, The Netherlands\\
}

\date{Accepted XXX. Received YYY; in original form ZZZ}

\pubyear{2015}

\begin{document}
\label{firstpage}
\pagerange{\pageref{firstpage}--\pageref{lastpage}}
\maketitle

\begin{abstract}
We report on the discovery of cool gas inflows towards three star-forming galaxies at $\left<z\right>\sim$ 2.30. Analysis of Keck Low-Resolution Imaging Spectrometer spectroscopy reveals redshifted low-ionisation interstellar (LIS) metal absorption lines with centroid velocities of 60 – 130 km s$^{-1}$. These inflows represent some of the most robust detections of inflowing gas into isolated, star-forming galaxies at high redshift. Our analysis suggests that the inflows are due to recycling metal-enriched gas from previous ejections. Comparisons between the galaxies with inflows and a larger parent sample of 131 objects indicate that galaxies with detected inflows may have higher specific star-formation rates (sSFR) and star-formation-rate surface densities (\SSFR). However, when additional galaxies without robustly detected inflows based on centroid velocity but whose LIS absorption line profiles indicate large red-wing velocities are considered, galaxies with inflows do not show unique properties relative to those lacking inflows. Additionally, we calculate the covering fraction of cool inflowing gas as a function of red-wing inflow velocity, finding an enhancement in high sSFR binned galaxies, likely due to an increase in the amount of recycling gas. Together, these results suggest that the low detection rate of galaxies with cool inflows is primarily related to the viewing angle rather than the physical properties of the galaxies.
\end{abstract}




\section{Introduction}

A key mechanism for the growth of galaxies is the conversion of cold gas into stars. However, the gas reservoir around galaxies can only sustain their star-formation rates (SFR) for a few gigayears \citep{Leroy08, Saintonge17, Tacconi18}. The accretion of pristine gas from the intergalactic medium (IGM) into galaxies is required to replenish their gas reservoirs throughout their evolution \citep{Kennicutt1983, Prochaska09, Bauermeister10}. The continuous accretion of metal-poor gas may also resolve the discrepancy between the number of low metallicity stars observed in the Milky Way and predictions from "closed-box" chemical evolution models, known as the G-Dwarf problem \citep{Bergh1962, Schmidt1963, Sommer-Larsen1991}. Additionally, cosmological simulations suggest that inflows of recycled gas from past outflows are the dominant source of accretion at $z$ < 1 \citep{Oppenheimer10, Henriques13, Alcazar17}. While inflows are commonly invoked to reconcile observations with theory, how gas accretes onto galaxies remains an open question, and is a top priority for astronomers in the coming decade \citep[see priority area: Unveiling the Drivers of Galaxy Growth,][]{Decadal}.

The problem of gas accretion from the IGM into galaxies has been studied extensively by simulations. In the ‘classical’ theory of galaxy formation, cold gas from the IGM falls into a dark matter potential well, shock heats to the virial temperature of the dark matter halo, forming a hot, gas-pressure-supported atmosphere in quasi-hydrostatic equilibrium, which can then radiatively cool and fall to the halo centre \citep{Rees1977, Silk1977, White1978}. More recently, a new theoretical paradigm emerged in which infalling cold gas does not shock heat to the virial temperature of the dark matter halo \citep[e.g.,][]{Keres05, Dekel06, Faucher11a}. In this picture, galaxies in dark matter halos below $\sim$$10^{12} M_{\odot}$ at any redshift freely accrete cold gas from the IGM via a “cold-mode”, as the gas cooling times are shorter than the dynamical times. Once inside the dark matter halo, large-scale tidal torques can align the accreting gas with a pre-existing disk, forming a warped, extended “cold-flow disk” that co-rotates with the central disk \citep[e.g.,][]{Stewart11}. Additionally, in massive halos ($M_{\rm{halo}}$ $\gtrsim 10^{12} M_{\odot}$) at $z \gtrsim$ 2, simulations predict that dense, collimated “cold-streams” of gas can penetrate the hot medium surrounding a massive galaxy and feed the central galaxy \citep[e.g.,][]{Dekel06, Dekel09}.

Although simulations suggest that the accretion of cold gas is crucial for the formation and evolution of galaxies, direct observational detections of such inflows are sparse. Fortuitously aligned galaxy–quasar pair studies indicate that gas traced by low-ionisation interstellar (LIS) metal absorption lines appears to co-rotate with the host galaxy, which is interpreted as evidence of a “cold-flow disk" \citep{Kacprzak10, Bouche13, Bouche16, Diamond16, Ho19, Zabul19}. However, there are different possible origins for the gas that cause LIS absorption lines, such as outflows that can carry angular momentum when launched from a rotating disk, complicating the tracing of accretion by co-rotating LIS absorption lines. Recently, studies have shown evidence of accretion from cold filamentary streams into the centre of massive haloes at $z$ $\sim$ 3 \citep{Daddi21, Fu21}. Other studies, using down-the-barrel observations, have reported redshifted LIS metal absorption lines in $\sim$5\% of star-forming galaxies at $z$ $\sim$ 1 \citep{Rubin12, Martin12}, similar to the small covering fractions predicted by simulations \citep[e.g.,][]{Faucher11b, Fumagalli11, Fumagalli14, Faucher15}. At higher redshifts, $z$ $\sim$ 2 -- 4.5, \cite{Calabro22} found redshifted LIS lines ($v$ > 0 km s$^{-1}$) in 34\% of their star-forming galaxies. However, the origin of inflowing gas in down-the-barrel observations is often ambiguous, as the gas must have moderate metallicity to give rise to the LIS absorption lines. The gas could be part of a filament of pristine, low-metallicity gas from the IGM that mixed with metal-enriched gas while moving through the circumgalactic medium (CGM) or the re-accretion of gas previously ejected from the galaxy. 

The low detection rate of inflowing gas may be due to (1) the geometry of accretion, and/or (2) weak redshifted absorption line profiles. As simulations and observations have shown, the covering fraction of inflows is relatively small \citep{Faucher11b, Faucher15}. To be observable, a filament would need to be well aligned with the line of sight so that a strong absorption feature is produced. Additionally, inflows may be missed altogether due to weak or absent absorption lines if the filament has low metallicity or a small velocity such that outflows or bulk ISM motions dominate the absorption line profiles \citep{Kimm11}.

In this paper, we report on the identification of three star-forming galaxies with observed inflows drawn from the MOSFIRE Deep Evolution Field \citep[MOSDEF;][]{MOSDEF} Survey, which have significant inflows based on LIS absorption lines measured from deep ($\sim$7.5 hrs) rest-UV observations from the Keck Low Resolution Imaging Spectrometer \citep[LRIS;][]{Oke1995, Steidel03}. The primary objectives of this study are to (1) explore which, if any, galactic properties differ between galaxies with robust inflows and outflows; and (2) measure the covering fraction of inflowing gas as a function of red-wing inflow velocity. The outline of this paper is as follows. Section \ref{sec:Data} describes the MOSDEF survey, follow-up LRIS spectroscopy, stellar population models, and the approach for measuring inflow velocity and galaxy properties. In Section \ref{sec:props}, we present our main results on the properties of the inflowing galaxies and discuss the implications in Section \ref{sec:discussion}. The conclusions are summarised in Section \ref{sec:conclusion}. Throughout this paper, we adopt a standard cosmology with $\Omega_{\Lambda}$ = 0.7, $\Omega_{M}$ = 0.3, and $\rm{H}_{0}$ = 70 km $\rm{s}^{-1} \rm{ Mpc}^{-1}$. All wavelengths are presented in the vacuum frame.

\section{Data and Measurements}
\label{sec:Data}

\subsection{MOSDEF Survey}

The galaxies presented in this paper were drawn from the MOSDEF Survey, which targeted $\approx$1500 $H$-band selected galaxies and AGNs at redshifts 1.4 $\leq$ $z$ $\leq$ 3.8. The survey obtained moderate-resolution ($R$ $\sim$ 3000$-$3600) near-infrared spectra using the Multi-Object Spectrometer for Infra-Red Exploration \citep[MOSFIRE;][]{McLean12} spectrograph over 48.5 nights between 2012 $-$ 2016. Galaxies were targeted for spectroscopy based on pre-existing spectroscopic, grism, or photometric redshifts that placed them in three redshift ranges ($z$ = 1.37 $−$ 1.70, $z$ = 2.09 $−$ 2.61, and $z$ = 2.95 $−$ 3.80) where strong rest-frame optical emission lines (e.g., \HB, [\ion{O}{III}], \HA, [\ion{N}{II}]) lie in the $YJHK$ transmission windows. For full details regarding the MOSDEF survey (targeting, data reduction, and sample properties), we refer readers to \cite{MOSDEF}.

Emission-line fluxes were measured by simultaneously fitting a line with the best-fit SED model for the continuum and a Gaussian function (see \cite{Reddy22} for a complete description of the SED modelling). For multiple lines that lie in close proximity, multiple Gaussians were fit, such as the [\ion{O}{II}] doublet and \HA\ and the [\ion{N}{II}] doublet, which were fitted with two and three Gaussians, respectively. Systemic redshifts were derived from the strongest emission line, usually \HA\ or [\ion{O}{III}]$\lambda$5008, and were used to fit the other rest-frame optical nebular emission lines. Further details on emission-line measurements and slit loss corrections are given in \cite{MOSDEF} and \cite{Reddy15}.

Galaxy sizes and inclinations were estimated from the effective radius (\RE), within which half the total light of the galaxy is contained, and the axis ratio ($b$/$a$), respectively, measured by \cite{vanderWel14}\footnote{\url{https://users.ugent.be/~avdrwel/research.html}} using GALFIT \citep{Peng10} on HST/F160W images from the CANDELS survey.

\subsection{MOSDEF-LRIS Spectroscopy}

\begin{figure*}
  \includegraphics[width=\linewidth, keepaspectratio] {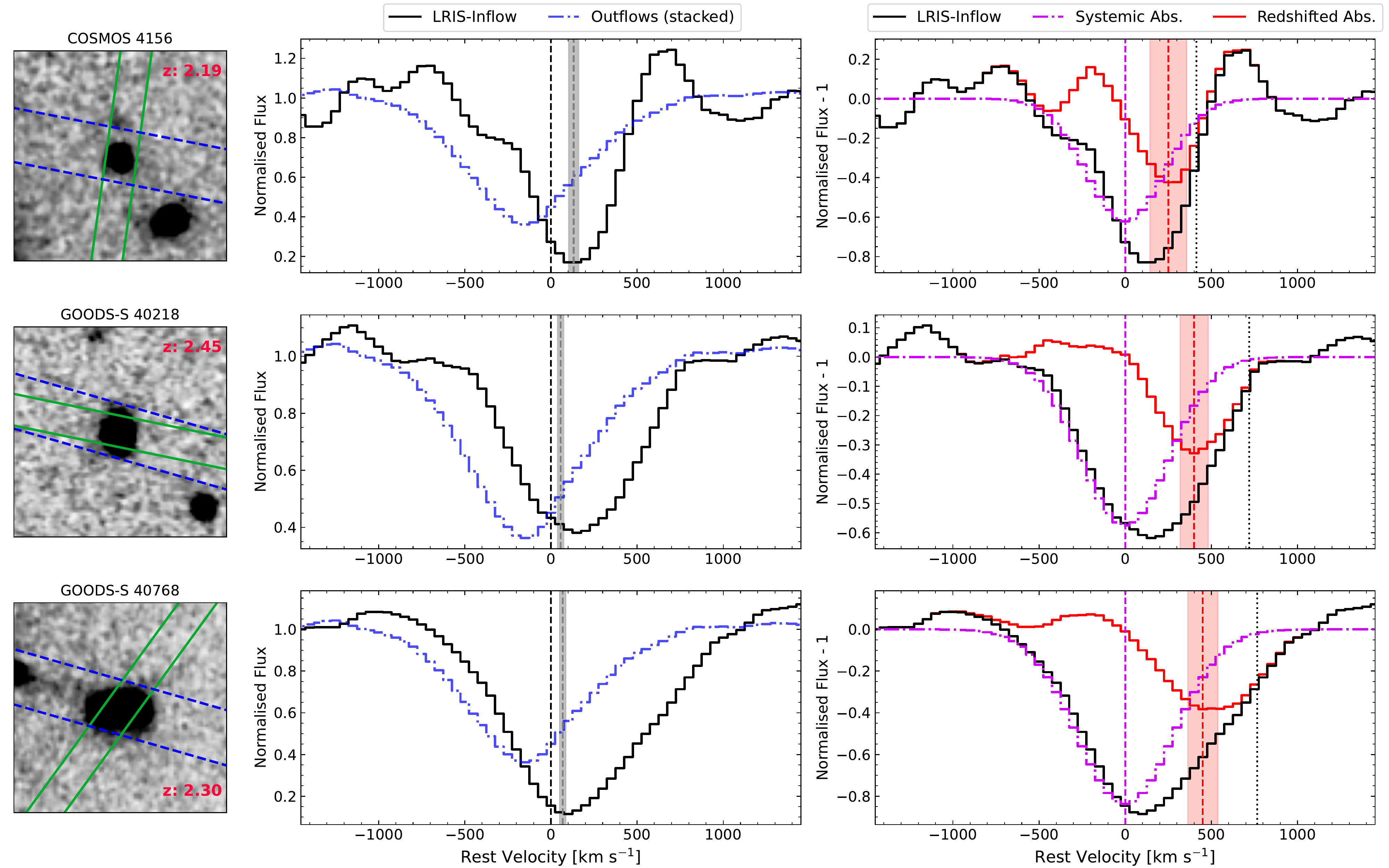}
  \vspace{-0.35cm}
  \caption{The three galaxies with significantly measured inflows. \textit{Left}: F160W HST images. Each image is 4\farcs8 on a side, which corresponds to an angular distance of 39 kpc at $z$ = 2.3, and oriented such that North is up and East is left. The solid green and dashed blue outlines represent the MOSDEF and LRIS slit placements, respectively. \textit{Centre}: Comparison between the composite LIS metal absorption line profiles of the three inflow galaxies and a stack of 29 outflow galaxies (\Vlis\ + 3$\sigma_{v_{\Delta_{\rm{LIS}}}}$ < 0 km s$^{-1}$) as solid black and blue dot-dashed lines, respectively. The centroid velocity (\Vlis) is marked with a dashed vertical line. The dark grey shaded region marks the 1$\sigma$ confidence interval. \textit{Right}: Decomposition of absoprtion line profile into symmetric, interstellar absorption (purple dashed Gaussian) and inflow absorption (solid red line) components, discussed in Section \ref{sec:decomp}. The centroid velocity of the inflow component is marked with a red dashed vertical line. The red shaded region marks the 1$\sigma$ confidence interval. For comparison, the dotted line marks the maximum red-wing velocity (see Section \ref{sec:vels}).
  }
  \label{fig:summary}
\end{figure*}

In this study, we use a subset of galaxies drawn from the MOSDEF survey with follow-up rest-frame UV spectroscopy, which provides coverage of absorption and emission features from  diffuse circumgalactic gas (e.g., \LyA, \ion{Si}{II}, \ion{O}{I}, \ion{C}{II}). Here, we present an overview of the sample and refer readers to \cite{topping2020} and \cite{Reddy22} for more details on the sample selection, data collection, and reduction. Objects for follow-up LRIS spectroscopy were prioritised based on strong detections of rest-optical emission lines (\HB, [\ion{O}{III}], \HA, and [\ion{N}{II}]), with higher priority given to objects with confirmed spectroscopic redshift at 1.90 $\leq$ $z$ $\leq$ 2.65. Additional objects were selected in the following order of priority: objects with \HA, \HB, and [\ion{O}{III}] detected at $\geq$ 3$\sigma$ and an upper limit on [\ion{N}{II}]; objects with a confirmed systemic redshift from MOSDEF; objects observed as part of the MOSDEF survey without a successful systemic redshift measurement, but with a prior spectroscopic or photometric redshift; and finally, objects not observed with MOSFIRE, but with a prior redshift from the 3D-HST survey that placed them within the redshift ranges and magnitude limit of the MOSDEF survey.

Observations were performed over nine nights in 2017 and 2018 in the COSMOS, GOODS-S, GOODS-N, and AEGIS fields using nine multi-object slit masks milled with 1\farcs2 slits. The instrumental setup included a dichroic to split the incoming beam at $\sim$5000\ \AA\ into the blue and red arms of LRIS. We configured the blue side with the 400 lines/mm grism, and the red side with the 600 lines/mm grating. This configuration provided continuous spectral coverage from the atmospheric cut-off at 3100\AA\ up to a typical wavelength of $\sim$7000\AA, depending on the position of the slit within the spectroscopic field of view. The seeing ranged from 0\farcs6 to 1\farcs2 with a typical value of 0\farcs8. The rest-frame spectra were continuum normalised around each LIS absorption line. The local continuum was determined by fitting a linear function between the average flux in two spectral windows, bluewards and redwards of the LIS absorption line. The spectral windows, listed in Table \ref{tbl:spectral_windows}, were chosen to bracket the line and exclude other spectral features.

\subsection{Galaxy Properties and Measurements}

\subsubsection{Sample Selection}
\label{sec:sample}

Several criteria were applied to the parent MOSDEF-LRIS sample to create a sample with robust systemic ($z_{\rm{sys}}$) and LIS absorption line redshifts ($z_{\rm{LIS}}$) for our analysis. We select objects with secure systemic redshifts from MOSDEF spectroscopy. Specifically, objects must have more than one emission line with an integrated line flux with S/N $\geq$ 2. Next, any objects for which the LRIS spectra contained irreparable artifacts that were too noisy to yield a robust absorption-line redshift were removed. Active galactic nuclei (AGNs) identified by IR colors, X-ray emission, and/or the [\ion{N}{II}]/\HA\ line ratio were removed \citep{Coil15, Azadi17, Azadi18, Leung19}. Finally, objects for which the MOSFIRE or LRIS spectra indicate that the target may be blended with a foreground object were removed. 

These criteria result in a final sample of 134 galaxies, of which 39 galaxies (29\%) exhibit redshifted LIS metal absorption lines. For our inflow analysis, galaxies that have redshifted absorption lines with centroids statistically consistent with being redshifted (\Vlis\ - 3$\sigma_{v_{\Delta_{\rm{LIS}}}}$ > 0 km s$^{-1}$, see Section \ref{sec:vels}) are considered to have inflowing gas, reducing the sample to three galaxies (hereafter the “LRIS-Inflow” sample). Similarly, 29 galaxies have blueshifted centroids (\Vlis\ + 3$\sigma_{v_{\Delta_{\rm{LIS}}}}$ < 0 km s$^{-1}$) indicating outflowing gas. Figure \ref{fig:summary} presents images, composite and decomposed (see Section \ref{sec:decomp}) LIS absorption line profiles of the LRIS-Inflow galaxies. Figure \ref{fig:all_spec} presents individual optical emission and LIS absorption lines of the LRIS-Inflow galaxies.

\begin{figure}
  \includegraphics[width=\columnwidth, keepaspectratio]{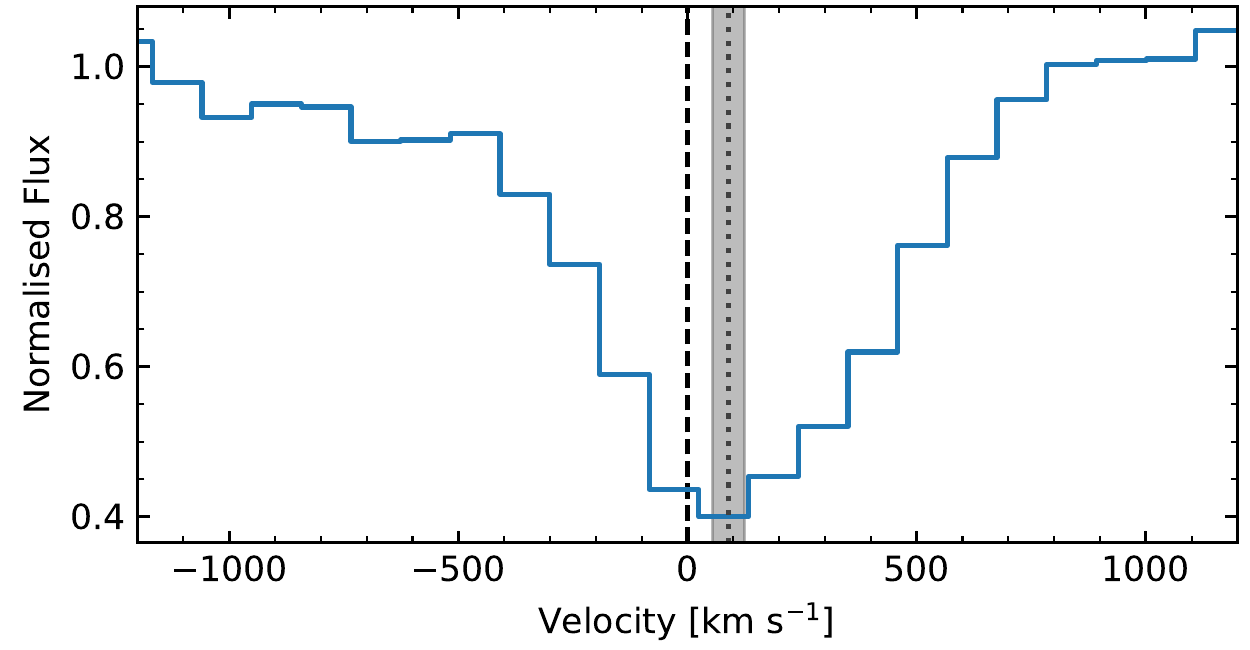}
  \vspace{-0.5cm}
  \caption{Average LIS metal absorption profile of GOODS-N 24328. An example of a galaxy with a negatively skewed profile (Skewness Ratio = -0.47$\pm$0.17) added to the LRIS-Inflow galaxies to create a larger "inflow" subsample. The centroid velocity (\Vlis) is marked with a dotted vertical line. The dark grey shaded region marks the 1$\sigma$ confidence on the centroid velocity.}
  \label{fig:subsample}
\end{figure}

As the absorption line centroids reflect the overall velocity distribution of the gas, galaxies may have a large amount of inflowing gas masked by more prevalent outflowing gas or gas at zero systemic velocity. To increase the sample size of "inflowing" galaxies, we consider galaxies whose LIS absorption line profiles are skewed redward, indicating a large fraction of inflowing gas, see Figure \ref{fig:subsample}. We define a skewness ratio \citep{Keerthi22}\footnote{The Skewness Ratio presented in this study measures the skewness of the absorption line profile with respect to 0 km s$^{-1}$, as oppose to \cite{Keerthi22} which measures skewness with respect to the centroid velocity.} as
\begin{align}
    \text{Skewness\ Ratio}  = \left| \frac{v_{\rm{max, blue}}}{v_{\rm{max, red}}} \right|-1,
\end{align}
where \Vmaxb\ and \Vmaxr\ are the outflow and inflow and velocity at 100\%  of the continuum (Section \ref{sec:vels}). A negative Skewness Ratio indicates that the red-wing is more extended than the blue-wing. Galaxies with Skewness Ratio + 2$\sigma_{\rm{Skewness\ Ratio}}$ < 0 and \Vmaxr\ $-$ \Vmaxb\ > 300 km s$^{-1}$, corresponding to the velocity resolution of the LRIS observations, are added to the LRIS-Inflow sample to create a second, larger sample of galaxies with inflowing gas. We adopt these thresholds because they are rather conservative, such that the selected galaxies likely have a substantial amount of inflowing gas.

\subsubsection{Velocity Measurements}
\label{sec:vels}

Using systemic redshifts and  LIS absorption line redshifts, we measured centroid velocities from the redshift difference:
\begin{equation}
\Delta v_{\rm{LIS}} = \frac{c(z_{\rm{LIS}} - z_{\rm{sys}})}{1 + z_{\rm{sys}}},
\end{equation}
where $z_{\rm{LIS}}$ is taken from \cite{topping2020}.  Briefly, $z_{\rm{LIS}}$ was obtained by fitting the absorption lines with Gaussian functions and a quadratic function for the local continuum, then determining the centroids of the Gaussians. Uncertainties were determined by perturbing the LRIS spectra by the corresponding error spectra, refitting the lines, and recalculating the centroids. Any LIS absorption lines with poor fits were excluded in the calculation of the average $z_{\rm{LIS}}$, which is typically based on two lines for a given galaxy. The uncertainties of the centroid velocities ($\sigma_{v_{\Delta_{\rm{LIS}}}}$) are taken as the standard deviation of many realisations after perturbing $z_{\rm{sys}}$ and $z_{\rm{LIS}}$ by their corresponding errors and recalculating \Vlis.

In addition to centroid velocities, another technique for estimating velocities uses the wings of the absorption line profile. In the down-barrel-observations, \Vlis\ represents the sum of all outflowing, inflowing, and interstellar gas at or near $z_{\rm{sys}}$. Gas near the systemic redshift will naturally shift the centroid towards zero systemic velocity. To better estimate the velocity of inflowing (outflowing) gas, we consider the red (blue) wings of the absorption line profiles. Previous outflow studies have either used the velocity where the absorption feature reaches some percent of the continuum level \citep{Martin05, Weiner09, Chisholm15, Du16, Weldon22} or the maximum velocity where the absorption feature returns to the continuum level \citep{Steidel10, Kornei12, Rubin14, Weldon22}. We measure the inflow and outflow velocity at 80\% (\Vfrcr, \Vfrcb) and 100\%  (\Vmaxr, \Vmaxb) of the continuum following a similar approach as \cite{Kornei12} and \cite{Weldon22}. Using the normalised spectra, we identify the absolute minimum of a detected absorption feature, then move towards longer ($v_{\rm{red}}$) or shorter ($v_{\rm{blue}}$) wavelengths, checking the sum of the flux and its uncertainty at each wavelength step. We record the first wavelengths at which this sum exceeds 0.8 and 1.0, for \Vfrc\ and \Vmax, respectively, perturb the spectrum by its error spectrum and repeat the same procedure many times. The average and standard deviation, after 3$\sigma$ clipping, of the trials are then used to calculate \Vfrc, \Vmax, and their uncertainties. This entire process was repeated for each detected LIS feature, listed in Table \ref{tbl:spectral_windows}, adopting \Vfrc\ and \Vmax\ as the average of the detected LIS features. For the 134 galaxies, \Vfrcr\ (\Vmaxr) ranges from $-$228 to 662 km s$^{-1}$ ($-$218 to 766 km s$^{-1}$) with a mean of 200$\pm$150 km s$^{-1}$ (300$\pm$188 km s$^{-1}$). For the LRIS-Inflow galaxies, \Vfrcr\ (\Vmaxr) ranges from 328 to 662 km s$^{-1}$ (413 to 766 km s$^{-1}$) with a mean of 525$\pm$143 km s$^{-1}$ (720$\pm$156 km s$^{-1}$).

\begin{table}
  \centering
  \caption{Spectral Windows}
  \label{tbl:spectral_windows}
  \begin{threeparttable}
    \begin{tabular}{lccc}
        \hline
        \hline
        Line  & $\lambda_{\rm{rest}}$ (\AA)\tnote{a} & Blue Window (\AA)\tnote{b} & Red Window (\AA)\tnote{b}\\
        \hline
   
        Ly$\alpha$   & 1215.67 & 1195 - 1202 & 1225 - 1235\\
        \ion{Si}{II} & 1260.42 & 1245 - 1252 & 1270 - 1275\\
        \ion{O}{II} + \ion{Si}{II} & 1303.27 & 1285 - 1293 & 1312 - 1318\\
        \ion{C}{II}  & 1334.53 & 1320 - 1330 & 1342 - 1351\\
        \ion{Si}{II} & 1526.71 & 1512 - 1520 & 1535 - 1540\\
    \hline
    \end{tabular}
    \begin{tablenotes}
        \item[a] Rest-frame vacuum wavelength, taken from the Atomic Spectra Database website of the National Institute of Standards and Technology (NIST), https://www.nist.gov/pml/atomic-spectra-database.
        \item[b] Wavelength window over which continuum fitting was performed.
    \end{tablenotes}

  \end{threeparttable}
\end{table}

\subsubsection{Decomposition into Symmetric and Redshifted Absorption}
\label{sec:decomp}

A complementary approach to isolate the inflowing gas is to separate its absorption from the intrinsic absorption of interstellar gas. In the down-barrel-observations, the LIS metal absorption lines probe the velocity of gas seen along the line of sight towards a galaxy. Interstellar gas within the galaxy will absorb close to the systemic redshift (i.e., $v \sim$ 0 km s$^{-1}$) with a velocity range set by the galactic rotation curve and velocity dispersion. Inflowing gas gives rise to redshifted absorption (at $v$ > 0 km s$^{-1}$), while outflowing gas produces blueshifted absorption ($v$ < 0 km s$^{-1}$). In principle, the redshifted (blueshifted) absorption is a mixture of interstellar and inflowing (outflowing) absorption. However, without detections of driven outflows from the LRIS-Inflow galaxies, it is difficult to separate outflowing gas from the blue-wing of interstellar absorption. To determine properties of the inflowing gas, we attempt to separate it from the systemic, interstellar gas by adopting a simple three-component model:

\begin{equation}
    F_{\rm{obs}} = C(\lambda) + A_{\rm{sys}} + A_{\rm{inflow}},
\end{equation}
where $F_{\rm{obs}}$ is the observed flux density, C($\lambda$) is the underlying linear continuum of the galaxy, $A_{\rm{sys}}$ is the systemic absorption from interstellar gas at or near $z_{\rm{sys}}$, and $A_{\rm{inflow}}$ is the absorption from inflowing gas.

We begin by creating a composite LIS absorption line profile. For each galaxy, the flux and error around each detected LIS line are interpolated onto a common velocity grid. The composite flux is then taken as the average flux of the detected lines, and the error is estimated by adding the error of the lines in quadrature. For the decomposition, we perform two preliminary fits and one final fit. The first preliminary fit uses a linear continuum with a Gaussian to fit the absorption profile using \texttt{curve fit}, a non-linear least squares fitting routine from the \texttt{scipy.optimize} subpackage. We use this fit to divide the composite spectra by a linear continuum, normalising the spectrum to 1. Next, we fit a Gaussian for the systemic absorption to the blue-side of $F_{\rm{obs}}$/C -- 1, using only the pixels at --1600 < v < 0 km s$^{-1}$. The free parameters are a constant continuum level, absorption intensity, and velocity dispersion; the central velocity is held fixed at 0 km s$^{-1}$. We use the resulting values and errors of this second fit as initial values for the final fit, which is done using \texttt{emcee}, a Python Markov chain Monte Carlo (MCMC) ensemble sampler. Finally, we subtract the systemic absorption from the normalised spectra to obtain the redshifted absorption of the inflowing gas.

The right panels of Figure \ref{fig:summary} show the decomposition applied to LRIS-Inflow galaxies. The centroid velocity of the inflowing gas (\Vcenr; red dashed line) is measured at the absolute minimum of the redshifted absorption component, and its error is estimated by perturbing the composite profile by its error, and repeating the decomposition many times.  The velocity of the inflow absorption presents an intermediate case between the centroid of the observed profile and maximum red-wing velocity.

\subsubsection{Galaxy properties}
\label{sec:gal_props}

In this study, we analyse several of the global galaxy properties (e.g., SFR, mass, star-formation-rate surface density, inclination) discussed in \cite{Weldon22}. Here, we briefly summarise the measurements and refer readers to \cite{Weldon22} for more details. Stellar masses (\Mste) and SFRs were derived from spectral energy distribution (SED) modelling, adopting a \citet[hereafter BC03]{BC03} stellar population synthesis model, \cite{Chabrier03} initial mass function, constant star formation histories (SFH), Small Magellanic Cloud (SMC) attenuation curve \citep{Fitzpatrick90, Gordon03}, and sub-solar metallicity ($Z_{\ast} = 0.28 Z_{\odot}$) \footnote{A steep SMC-like attenuation curve and sub-solar metallicities have been found to provide self-consistent SFRs with those derived using other methods \citep{Reddy18b, Reddy22}. However, other studies have suggested that a \cite{Calzetti00} attenuation curve and solar metallicity provide a better description for high-mass (log($M_{\star}/M_{\odot}$) $\ge$ 10.04) star-forming galaxies at $z \sim$ 2 \citep{Reddy18a, Shivaei20}. If instead we assume a \cite{Calzetti00} attenuation curve and solar metallicities for high-mass galaxies, on average, stellar masses are 0.06 dex higher and SFRs are 0.4 dex higher. Our main results do not significantly change if we were to alter the assumed attenuation curve.}. The sample has a stellar mass range of 8.6 < log($M_{\star}$/$M_{\odot}$) < 10.9 with a median log($M_{\star}$/$M_{\odot}$) of 9.9 and SFR range from 0.32 < log(SFR/$M_{\odot}$ yr$^{-1}$) < 1.97 with a median log(SFR/$M_{\odot}$ yr$^{-1}$) of 0.93. In addition, dynamical (\Mdyn) and baryonic mass (\Mbar\ = \Mdyn\ + \Mgas) were calculated following the procedure of \cite{Price20}. Dynamical masses were derived using circular velocities measured from 2D spectra with detected rotation or inferred using integrated velocity dispersions and the best-fit ensemble $V/\sigma$ from galaxies without detected rotation, while \Mgas\ was estimated from the Schmidt-Kennicutt \citep{Kennicutt89} relation between $\rm{\Sigma_{SFR}}$ and $\rm{\Sigma_{gas}}$. 

Absorption-line studies indicate that outflows are ubiquitous in $z >$ 2 star-forming galaxies \citep[e.g.,][]{Shapley03, Steidel10}. Observations suggest that outflow velocity increases with the SFR and star-formation-rate surface density (\SSFR) of a galaxy \citep[e.g., see][and references therein]{Weldon22}. This result implies that detectable inflowing gas may only occur in galaxies with low SFRs or \SSFR\ when outflows would be weak or absent. We have chosen to focus on SFR[SED] when discussing SFR as of the 134 galaxies 23\% lack significant \HA\ and/or \HB\ detections. However, as the SFR[SED] is tightly correlated with stellar mass (i.e., both quantities are sensitive to the normalisation of the best-fit SED), \HA\ SFRs are used to calculate specific SFR (sSFR), using the conversion factor from \cite{Reddy18b} for a BC03 stellar population synthesis model and sub-solar metallicity adopted for the SED fitting. The star-formation-rate surface density is then defined as \SSFR\ = SFR[SED]/(2$\pi R_{\rm{E}}^{2}$). Additionally, at a given SFR, outflows may be more effectively launched from a shallow galaxy potential (i.e., low stellar, dynamical, and/or baryonic mass) relative to a deep potential \citep{Reddy22}. To examine the frequency of galaxies with observed inflows on both \SSFR\ and the galaxy potential, we define the specific star-formation-rate surface density (\sSSFR) as \sSSFR\ = SFR[\HA]/(2$\pi R_{\rm{E}}^{2} M_{X}$), where $M_{X}$ is the stellar, dynamical, or baryonic mass.

\begin{figure}
  \includegraphics[width=\columnwidth, keepaspectratio]{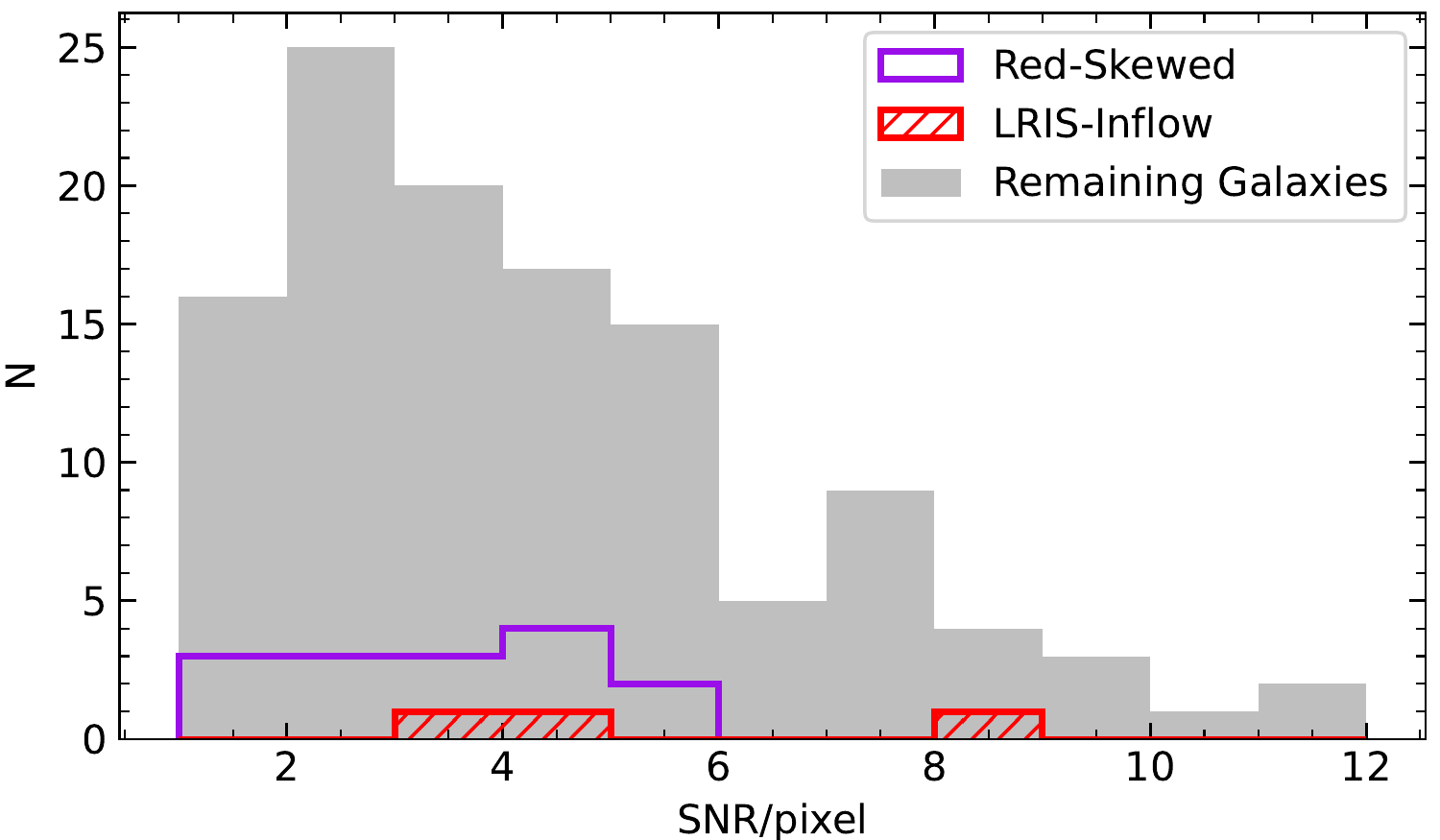}
  \vspace{-0.35cm}
  \caption{Distribution of UV continuum signal-to-noise ratio (SNR) per pixel over the wavelength range 1425\AA\ $\leq \lambda \leq$ 1500\AA. Hashed red, open purple, and solid grey bars represent inflowing, redward-skewed, and remaining galaxies, respectively. Neither the LRIS-Inflow nor red-skewed galaxies appear biased towards higher SNRs, suggesting that the SNR does not play a significant role in whether their properties differ from the remaining galaxies.}
  \label{fig:SNR}
\end{figure}

\begin{figure*}
  \includegraphics[width=\linewidth, keepaspectratio]{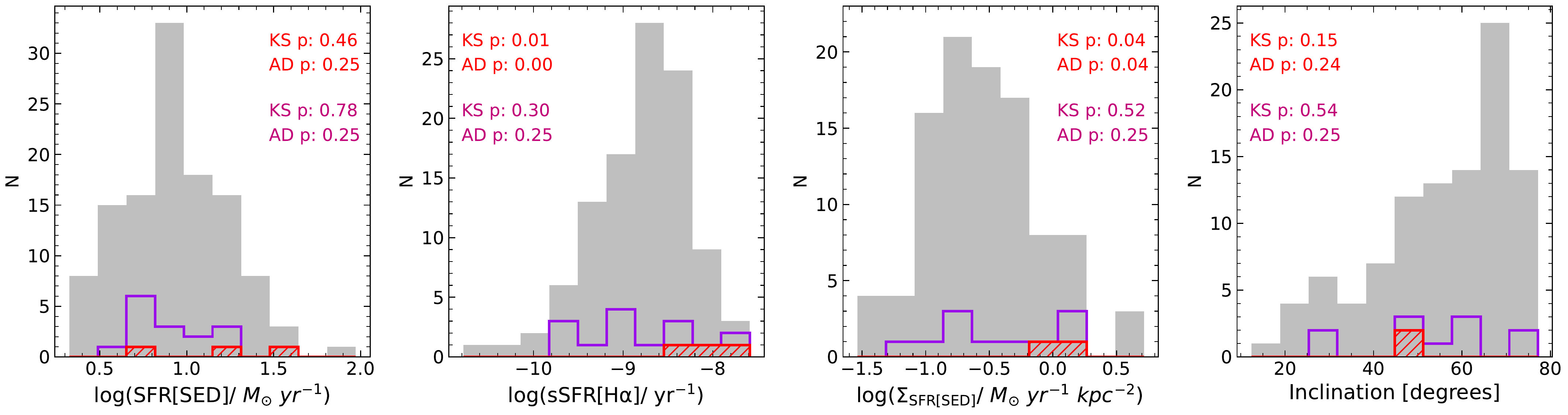}
  \vspace{-0.45cm}
  \caption{The distribution of various galactic properties, with the same style as Figure \ref{fig:SNR}. \textit{left}: log(SFR), \textit{left centre}: log(sSFR), \textit{right centre}: log($\Sigma_{\rm{SFR}}$), \textit{right}: Inclination. The p-values of the KS and AD tests between the inflowing and remaining galaxies (red) and inflowing, redward-skewed and remaining galaxies (magenta) are shown in the upper corners of each panel. Of the three inflowing galaxies, one does not have a robust \RE\ limiting our analysis to two inflowing galaxies for \SSFR\ and inclination.}
  \label{fig:prop_dist}
\end{figure*}

\section{Properties of Inflowing Galaxies}
\label{sec:props}

In this section, we compare the properties of the three LRIS-Inflow galaxies to those of the remaining LRIS galaxies. Of particular interest is whether the star-formation activity or inclination of the two groups differ significantly. In typical star-forming galaxies, inflows are theorised to enter along the major axis, either as part of a filament or "cold-flow disk", while outflows emerge perpendicular to the disk in a biconical structure \citep[e.g.,][]{Katz93, Bordoloi11, Stewart11, Bouche12, Kacprzak12}. On the other hand, outflows are theorised to be driven by energy injected into the ISM by supernovae; radiation pressure acting on cool, dusty material; cosmic rays; or a combination of these mechanisms \citep{Ipavich75, Chevalier85, Murray05, Murray11}. As galactic outflows are a common feature of actively star-forming galaxies \citep[e.g.,][]{Shapley03, Steidel10, Weldon22}, observable inflows may be biased towards edge-on galaxies with low star-formation rates, where inflows could potentially dominate over outflows. 

The detection of LIS metal absorption lines shifted away from systemic redshift requires a high signal-to-noise ratio (SNR) in the UV continuum. However, galaxies with high SNR are also associated with a high SFR. Therefore, it is important to check whether the three LRIS-inflow galaxies may be biased towards higher SFRs relative to the underlying sample. To investigate this possibility, we calculate the SNR per pixel over the wavelength range 1425\AA\ $\leq \lambda \leq$ 1500\AA. As shown in Figure \ref{fig:SNR}, neither the LRIS-Inflow (red-hashed) nor the red-skewed (open purple) galaxies are biased toward high SNRs. The SNR of the continuum does not appear to play a significant role in whether the properties of these subsamples differ from the remaining galaxies.

For each property (e.g., SFR, mass, \SSFR, inclination), we investigate whether the LRIS-Inflow galaxies have unique properties compared to the remaining galaxies. As shown in Figure \ref{fig:prop_dist}, the LRIS-Inflow galaxies appear to have higher sSFR and \SSFR\ relative to the full sample, suggesting that both SFR and the gravitational potential well of a galaxy may be important factors in determining the visibility of inflows. To quantitatively test whether the LRIS-Inflow galaxies are drawn from the same parent distribution as the remaining galaxies, we perform a Kolmogorov–Smirnov (KS) and Anderson-Darling (AD) test. While these statistical tests are similar, the AD test is more sensitive to the tails of the distributions than the KS test, which makes it a more powerful statistic when dealing with small sample sizes \citep{Hou09}.

Of the properties investigated, both the KS and AD tests indicate a $<$5\% probability that the sSFR and \SSFR\ of the LRIS-Inflow and remaining galaxies are drawn from the same distribution. On the other hand, the LRIS-Inflow galaxies have moderate ($\sim$45$^{\circ}$) inclinations. In part, these inclinations may reflect the lack of thin disks and/or the difficulty of measuring structural properties robustly for high redshift galaxies. The simple picture of biconical outflows along the minor axis and inflows along the major axis may not be applicable for galaxies without established disks. Rather, such galaxies may be fed primarily by filamentary inflows that enter at random angles, resulting in no significant relation between galaxy inclination and the detection of inflows. However, we caution that these results are based on a small number of galaxies\footnote{One LRIS-Inflow galaxy does not have a robust \RE\ limiting our analysis to two inflowing galaxies for \SSFR\ and inclination.}. 

To extend this analysis towards larger samples, we perform KS and AD tests on the distribution of properties between the inflowing galaxies, redward-skewed galaxies (Section \ref{sec:sample}) and the remaining galaxies. The tests indicate that all of the properties investigated are statistically consistent with having been drawn from the same parent distributions. Given the lack of significantly different properties with this larger subsample, we suggest that the detection of inflowing gas is likely dependent on the geometry and covering fraction of the inflowing gas into a host galaxy’s potential well rather than the particular physical properties of the host galaxy. In the next section, we explore the accretion geometry of the galaxies.

\begin{table*}
  \centering
  \caption{Inflow Velocities}
  \label{tbl:halo_vel}
  \begin{threeparttable}
    \begin{tabular}{llccccc}
        \hline
        \hline
        Field & V4ID & log($M_{\rm{halo}}/M_{\odot}$) & log$(M_{\star}/M_{\odot})$ & $v_{\rm{cen, red}}$\tnote{a} & $v_{\rm{max, red}}$\tnote{b} & $V_{\rm{stream}}$\tnote{c}\\
        & & & & [km s$^{-1}$] & [km s$^{-1}$] & [km s$^{-1}$]\\
        \hline
        COSMOS  & 4156  & 11.53$\pm$0.07 &  9.11$\pm$0.04 & 250$\pm$100 & 413$\pm$51 &  98$\pm$11\\
        GOODS-S & 40218 & 11.90$\pm$0.09 &  9.79$\pm$0.04 & 400$\pm$80  & 720$\pm$36 & 145$\pm$13\\
        GOODS-S & 40768 & 12.06$\pm$0.06 & 10.09$\pm$0.01 & 450$\pm$87  & 766$\pm$61 & 158$\pm$11\\
        \hline

    \end{tabular}
    \begin{tablenotes}
        \item[a] Centroid velocity of the inflowing gas (See Section \ref{sec:decomp}).
        \item[b] Maximum inflow velocity (See Section \ref{sec:vels}).
        \item[c] Cold stream inflow velocity
    \end{tablenotes}

  \end{threeparttable}
\end{table*}

\section{Discussion}
\label{sec:discussion}

\subsection{Comparison to Simulations}
\label{sec:simulations}

We report on three star-forming galaxies at $\left<z\right>\sim$ 2.3, whose spectra show redshifted LIS metal absorption lines. As these detections were made using LIS metal lines, they likely trace relatively metal-enriched gas rather than pristine gas accreting from the IGM for the first-time. The origin of this gas could be the re-accretion of gas previously ejected from the galaxy or gas-rich satellite dwarf galaxies being stripped and accreted onto the central galaxy. Several cosmological simulations have investigated the origin of gas accreted onto galaxies by considering pristine inflows, recycling gas, and interactions with satellite galaxies \citep[e.g.,][]{Alcazar17, Grand19, Mitchell20}. While simulations disagree on the relative contributions to total gas accretion, they have found that pristine inflows \citep{Mitchell20} or gas recycling \citep{Alcazar17} dominate the total accretion at $z\sim$2, with satellite mergers and stripping contributing a non-trivial but minor amount. Without evidence of dwarf satellite galaxies in HST imaging, we adopt the interpretation that the redshifted LIS lines are evidence of the re-accretion of gas previously ejected from the galaxy. However, the LIS absorption lines may also arise from filamentary inflows from the IGM. Several studies have shown that the circumgalactic medium (CGM) has a complex, multiphase structure in which metal-enriched gas may be distributed throughout \citep[e.g.,][]{Tumlinson17, Pointon19}. As pristine gas from the IGM moves through the CGM, it may mix with enriched gas before accreting onto the galactic disk, thus giving rise to the LIS absorption lines \citep{Faucher15}. However, the efficiency of mixing in the CGM remains highly uncertain and requires high-resolution simulations to properly resolve the small-scales where this mixing would take place. 

To investigate the origin of the inflowing gas in the LRIS-Inflow galaxies, we compared their redshifted centroid velocities and maximum red-wing inflow velocities to their predicted circular halo velocities ($V_{\rm{circ,halo}}$). Cosmological simulations suggest that the average radial inflow velocity of filamentary streams is between 0.5$V_{\rm{circ,halo}}$ and 0.8$V_{\rm{circ,halo}}$ \citep[e.g.,][]{Keres05, Goerdt15}. The $V_{\rm{circ,halo}}$ is calculated using the following equations from \cite{Mo02}:
\begin{equation}
    V_{\rm{circ,halo}} = \left( \frac{GM_{\rm{halo}}}{r_{\rm{halo}}}  \right)^{1/2}
\end{equation}
\begin{equation}
    r_{\rm{halo}} = \left( \frac{GM_{\rm{halo}}}{100\Omega_{m}H_{0}^{2}} \right)^{1/3} \left( 1+z \right)^{-1},
\end{equation}
where $M_{\rm{halo}}$ is the inferred halo mass from the redshift-dependent stellar-halo mass ratio from \cite{Behroozi19}. The conversion factor between $V_{\rm{circ,halo}}$ and filamentary stream inflow velocity ($V_{\rm{stream}}$) is calculated using the redshift-$M_{\rm{halo}}$ dependent function from \cite{Goerdt15}.

As listed in Table \ref{tbl:halo_vel}, the LRIS-Inflow galaxies have \Vcenr\ ranging from 250 to 450 km s$^{-1}$ and \Vmaxr\ ranging from 410 to 770 km s$^{-1}$, well above the 98 to 158 km s$^{-1}$ predictions for accretion of pristine gas from filamentary streams or the typical velocity dispersion ($\sim$80 km s$^{-1}$) of galaxies in the MOSDEF sample \citep{Price20}. These large inflow velocities imply that the LIS metal lines are tracing motion separate from filamentary inflows or large-scale ISM motion. Taken together, the large inflow velocities and metal-enrichment of the gas that gives rise to the redshifted LIS lines suggest that these lines are likely tracing the re-accretion of metal-enriched gas previously ejected from the galaxy.

\subsubsection{Future Simulations}

Simulations have focused on measuring flow properties, such as outflow rates and mass-loading factors, that depend on the geometry of the flows, which makes them notoriously difficult to compare with observations.However, a different approach is to directly compare observed absorption line profiles to "mock spectra" generated from simulations. As the spatial distribution of the gas is known, simulations could separate the outflowing, systemic, inflowing, and recycling gas components and assess their relative contributions to the observed absorption line profile. Tools such as TRIDENT \citep{Hummels17}, FOGGIE \citep{Peeples19}, and SALT \citep{Carr22} are promising for such future analyses. 

\subsection{Covering Fraction of Inflowing Gas}
\label{sec:covering_frac}

\begin{figure}
  \includegraphics[width=\columnwidth, keepaspectratio]{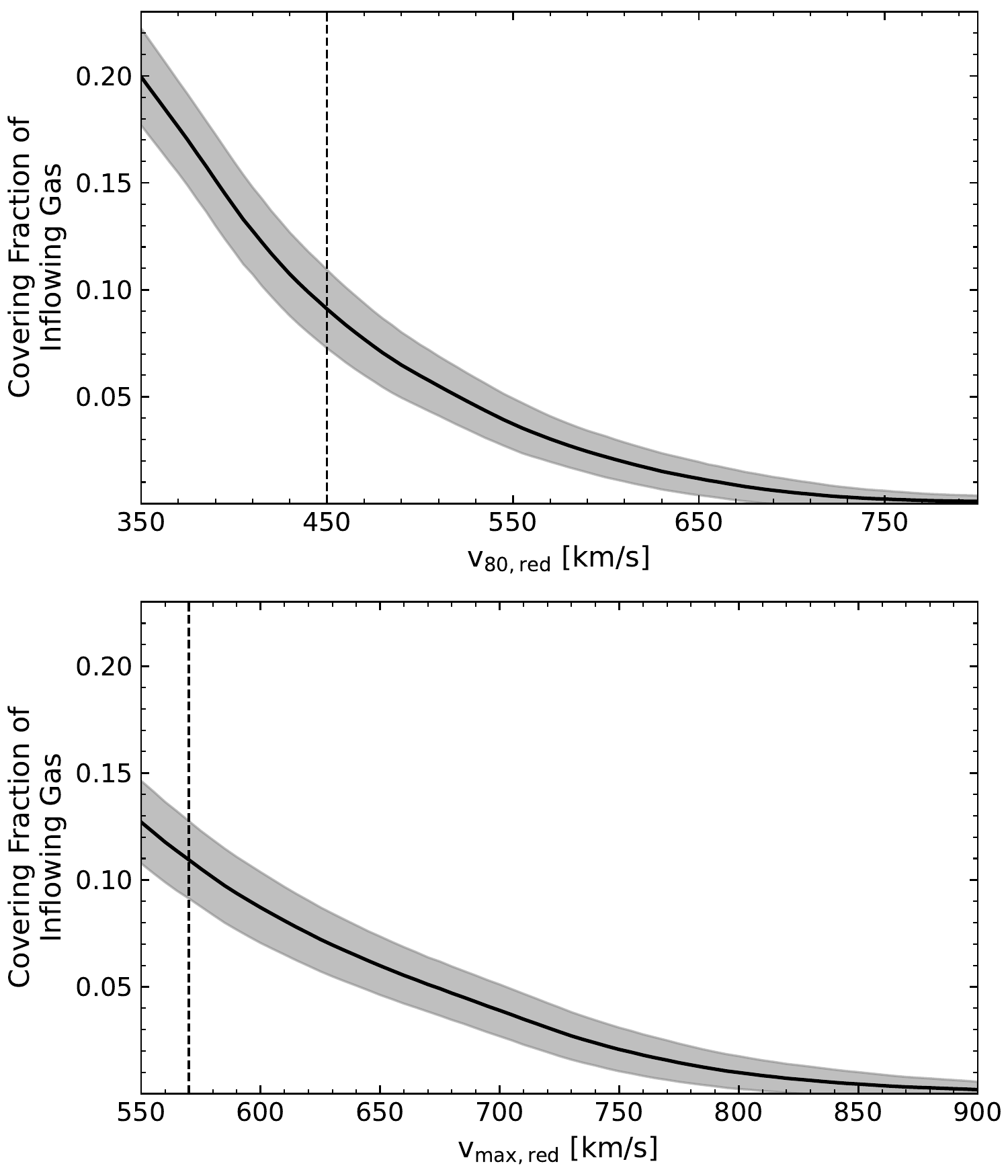}
  \vspace{-0.35cm}
  \caption{\textit{Top}: Covering fraction of inflowing gas versus \Vfrcr. \textit{Bottom}: Covering fraction of inflowing gas versus \Vmaxr. The covering fraction is taken as the fraction of galaxies with $v_{\rm{red}}$ equal to or greater than the given velocity. The grey shaded regions mark the 68\% confidence interval. The dashed vertical line marks the dividing velocity calculated using a subsample of galaxies with robust red-wing velocities ($v_{\rm{red}}$ $-$ 3$\sigma_{v_{\rm{red}}}$ > 0 km s$^{-1}$).}
  \label{fig:total_covering}
\end{figure}

\begin{figure*}
  \includegraphics[width=\linewidth, keepaspectratio]{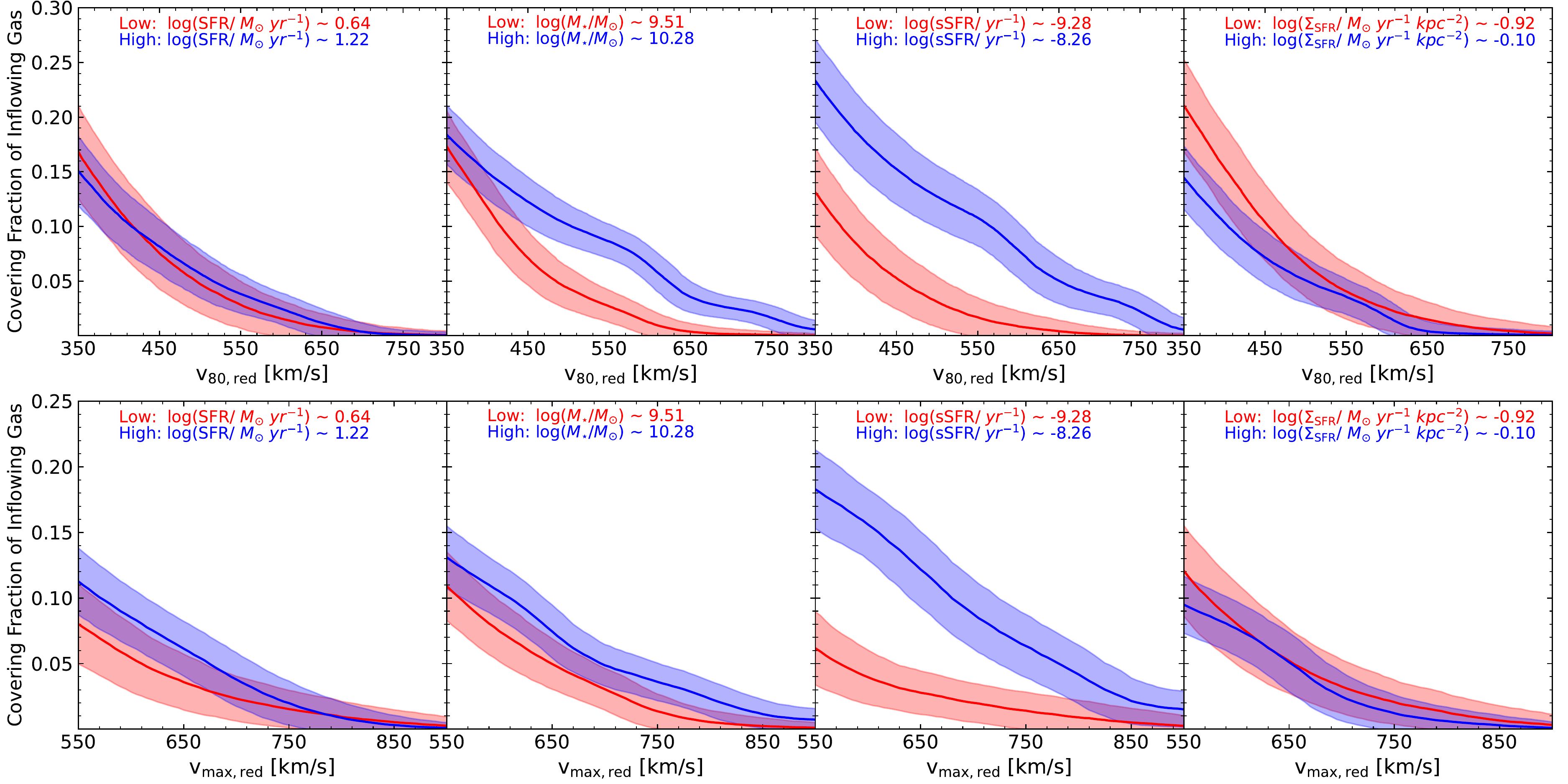}
  \vspace{-0.45cm}
  \caption{The covering fraction of inflowing gas in "lower" (red) and "higher" (blue) bins of \textit{left}: log(SFR), \textit{left centre}: log($M_{\star}/M_{\odot}$), \textit{right centre}: log(sSFR), and \textit{right}: log($\Sigma_{\rm{SFR}}$). The average value of each bin is listed in the top of each panel. Top panels are plotted against \Vfrcr\, and bottom panels are against \Vmaxr. The shaded regions mark the 68\% confidence interval.
  }
  \label{fig:binned_covering}
\end{figure*}

The LRIS-Inflow galaxies do not exhibit unusual properties compared to other galaxies, suggesting that the frequency of galaxies with observed inflows is related to the geometry and covering fraction of inflowing gas. Specifically, three galaxies with robust inflow velocities in a sample of 134 imply that along a random sightline through a galaxy, the chance of encountering an inflow (i.e., covering fraction) is $\sim$2\% $\pm$ 1\%. The low detection rate of inflowing gas in the sample is consistent with previous studies at intermediate redshifts. \cite{Rubin12} traced gas flows using Mg II and Fe II absorption lines in 101 $z \sim$ 0.5 star-forming galaxies of a similar SFR range but smaller stellar mass range compared to our sample (SFRs $\sim$ 1 – 63 $M_{\odot}$ yr$^{-1}$ and log($M_{\star}/M_{\odot}$) $\sim$ 9.5 - 11). They found six galaxies with redshifted absorption lines indicating inflowing gas at a 2$\sigma$ level. When compared to the remaining galaxies, their inflowing galaxies were highly inclined, suggesting that inflows are more likely along the major axis of a galaxy. Similarly, in a sample of 208 star-forming galaxies with SFRs $\sim$ 1 – 98 $M_{\odot}$ yr$^{-1}$ and stellar masses log($M_{\star}/M_{\odot}$) $\sim$ 9.4 – 11.5 at $z \sim$ 1, \cite{Martin12} detected inflowing gas at a 3$\sigma$ level in $\sim$4\% of galaxies. Both of these studies concluded that the low detection rate of inflowing gas was due to the low covering fraction of cold streams or recycled gas circulating in a galactic fountain. Alternatively, \cite{Calabro22} analysed 330 $z$ $\sim$ 2 -- 4.5 star-forming galaxies over a broad range of SFRs from 1 to 500 $M_{\odot}$ yr$^{-1}$, and stellar masses log($M_{\star}/M_{\odot}$) $\sim$ 8 – 10, finding galaxies with redshifted LIS lines (\Vlis\ $\geq$ 0 km s$^{-1}$) in 34\% of their sample. The authors suggest that their high detection rate of inflowing gas may be due to the increased role of inflows at earlier cosmic times. Although, we note that the higher detection rate of \cite{Calabro22} may be a matter of definition. In the MOSDEF-LRIS sample, 29\% of galaxies have \Vlis\ $\geq$ 0 km s$^{-1}$, similar to \cite{Calabro22}.

Using the full statistical power of the MOSDEF-LRIS sample, we investigated the covering fraction of inflowing gas as a function of red-wing inflow velocity. At a given velocity ($v$), we calculated the fraction of galaxies with red-wing velocities (\Vfrcr\ or \Vmaxr) equal to or greater than $v$, perturbed the velocities by their uncertainties, and repeated this calculation many times. The average and standard deviation of the fraction found for each trial were taken as the covering fraction and its uncertainty at each $v$. This calculation, however, is only meaningful for velocities above which inflows become prominent. Classifying galaxies as "inflowing" or "outflowing" based solely on their red-wing velocity is difficult as different combinations of centroids and line widths can produce the same absorption line wing (i.e., a narrow, redshifted line and a broad, blueshifted line could both have \Vmaxr\ = 100 km s$^{-1}$)\footnote{We note that red-wing velocity is also dependent on the spectral resolution of the observations.}. Due to this ambiguity, we defined a "dividing" velocity as the $v$ that maximised the fraction of galaxies with $v_{\rm{red}}$ > $v$ and \Vlis\ > 0 km s$^{-1}$ (true positives) or with $v_{\rm{red}}$ < $v$ and \Vlis\ < 0 km s$^{-1}$ (true negatives). At each $v$, we calculated the fraction of true positives and negatives, perturbed the red-wing velocities by their uncertainties, and repeated this many times. Figure \ref{fig:total_covering} shows the chance of encountering inflowing gas of at least speed $v$ along a random sightline through a galaxy (i.e., covering fraction) as a function of \Vfrcr\ and \Vmaxr. The covering fraction is roughly 20\%, 4\%, and 0.3\% for \Vfrcr\ larger than 350 km s$^{-1}$, 550 km s$^{-1}$, and 750 km s$^{-1}$, respectively. In contrast, the covering fraction is 12\% (2\%) for galaxies with \Vmaxr\ larger than 550 km s$^{-1}$ (750 km s$^{-1}$). The rapid decrease of inflowing gas covering fraction with red-wing velocity suggests that the low detection rate of galaxies with cool inflows is related to the viewing angle. As inflow velocity reaches speeds that may be more easily detected, the covering faction decreases to levels such that observing inflowing gas would be rare.

A complementary question is how, if at all, the inflow covering fraction varies with galactic properties. To investigate this, we divided the full sample into “lower” ($x$ + $\sigma_{x}$ < $x_{\rm{median}}$) and “higher” ($x$ - $\sigma_{x}$ > $x_{\rm{median}}$) property bins. However, this simple division for star-formation properties introduces significant SNR differences between the two bins. To account for the effect of continuum SNR, we remeasure $v_{\rm{red}}$ and $v_{\rm{blue}}$ for galaxies in the “higher” bins after adding random noise to their spectra, so that their SNR falls within the SNR range of galaxies in the “lower” bins. In Figure \ref{fig:binned_covering}, we plot the inflow covering fraction in bins of several galactic properties. The covering fraction appears independent of SFR, stellar mass, and \SSFR. On the other hand, in the “higher” sSFR bin, the covering fraction is enhanced by a factor of 1.8$\pm$0.6 (3.0$\pm$1.2) at the \Vfrcr\ (\Vmaxr) dividing velocity. It is not surprising that the galaxies with higher sSFR have an increased inflow covering fraction, as higher gas accretion would allow for more star formation per unit mass. The covering fraction could increase due to thicker filaments from the IGM or more recycling gas in the CGM from previous outflows. The down-the-barrel LRIS observations cannot distinguish between these cases. However, in either case, the LIS absorption line profiles would be red-skewed, as both cases increase the fraction of inflowing gas relative to outflowing and interstellar gas. As a test, we compared the skewness-ratio of the LIS absorption line profiles between the "lower" and "higher" bins. A KS-test indicates a 3\% probability that galaxies in the lower and higher sSFR bins are drawn from the same parent distribution, with the "higher" bin having a smaller skewness-ratio (i.e., more red-skewed) than the “lower” bin. Conversely, the SFR, stellar mass, and \SSFR\ bins each have a >40\% probability of being drawn from the same skewness distribution. Thus, the increased inflow covering fraction is likely due to more inflowing gas, such as recycling gas, which in turns increases the star-formation rate per unit mass.

\section{Conclusions}
\label{sec:conclusion}

We report on three star-forming galaxies from the MOSDEF Survey with additional deep rest-UV observations from Keck/LRIS with significantly measured centroid inflow velocities traced by LIS absorption lines. These inflows represent some of the most robust detections of inflowing gas into isolated, star-forming galaxies at $\left<z\right>\sim$ 2.3. Centroid velocities are measured from the redshift difference between $z_{\rm{sys}}$ and $z_{\rm{LIS}}$, while fractional (\Vfrc) and maximum (\Vmax) inflow (outflow) velocities are measured from the red (blue) wings of LIS lines that may better trace inflowing (outflowing) gas. Our main conclusions are as follows:

\begin{description}[leftmargin =1em]
    \setlength\itemsep{0.75em}
    \item[$\bullet$] The LRIS-Inflow galaxies have higher sSFR and \SSFR\ compared to the remaining galaxies, suggesting that both SFR and the gravitational potential of a galaxy are important in gas accretion. However, when other galaxies with large amounts of inflowing gas are included, no property is unique. The frequency of galaxies with observed inflows is then likely related to the geometry and covering fraction of inflowing gas (Section \ref{sec:props}). 
    \item[$\bullet$] The inflow centroid (\Vcenr) and maximum inflow velocities (\Vmaxr) of the LRIS-inflow galaxies are larger than predictions for the accretion of pristine gas from filamentary streams. We interpret the redshifted LIS absorptions lines of the LRIS-Inflow galaxies as tracing metal-enriched inflowing gas, such as recycled gas from previous ejections (Section \ref{sec:simulations}).
    \item[$\bullet$] At a conservative level, the detection of three galaxies with significant inflows in a sample of 134 implies a covering fraction of $\approx$2\% $\pm$ 1\%. Based on the full statistical power of the sample, the maximum covering fraction of cool inflowing gas at \Vfrcr\ = 350 km s$^{-1}$ is 20\% and at \Vmaxr\ = 550 km s$^{-1}$ is 12\%.
    \item[$\bullet$] Galaxies with higher sSFR have an increased inflow covering fraction, relative to those with lower sSFR. The larger covering fraction may be due to thicker filaments from the IGM or an increase in the amount of recycling gas in the CGM (Section \ref{sec:covering_frac}).    
\end{description}

Inflows of pristine gas from the IGM are required for galaxies to sustain their SFRs throughout their evolution. Here, we have presented three galaxies with significant inflows in a large sample of $z$ $\sim$ 2 galaxies that push the limits of current ground-based facilities, with full night ($\sim$7.5 hrs) observations needed to obtain sufficiently high SNR spectra. These spectra, however, can only give a glimpse into the complex nature of filamentary inflows and recycling gas. To build a better understanding of inflows, higher resolution spectroscopic data and multiple sightlines through the CGM are necessary to constrain the frequency and geometry of inflows and outflows around individual galaxies. The increased sensitivity of the next generation of 30-m extremely large telescopes will enable observations of faint background galaxies, increasing the density of sightlines through the CGM. In combination with deep IFU spectroscopy, studies will be able to probe the distribution and kinematics of cool gas throughout the CGM, which may allow one to differentiate between pristine gas accretion and enriched recycled material.

\section*{Acknowledgements}

We thank the anonymous referee for providing constructive feedback that improved the paper. We acknowledge support from NSF AAG grants AST1312780, 1312547, 1312764, and 1313171, grant AR13907 from the Space Telescope Science Institute, and grant NNX16AF54G from the NASA ADAP program. We thank the 3D-HST Collaboration, which provided the spectroscopic and photometric catalogs used to select the MOSDEF targets and derive stellar population parameters. This research made use of Astropy,\footnote{\url{http://www.astropy.org}} a community-developed core Python package for Astronomy \citep{Astropy13, Astropy18}. We wish to extend special thanks to those of Hawaiian ancestry on whose sacred mountain we are privileged to be guests. Without their generous hospitality, most of the observations presented herein would not have been possible.

\section*{Data Availability}

The MOSDEF data used in this article is publicly available and can be obtained at \url{http://mosdef.astro.berkeley.edu/for-scientists/data-releases/}. The LRIS data used in this article is available upon request.



\bibliographystyle{mnras}
\bibliography{ref} 




\appendix

\section{MOSFIRE and LRIS spectra}

Here we present plots of the MOSFIRE and LRIS spectra of the three galaxies with detected inflows. In each panel, the top row shows strong rest-optical emission lines, while the bottom row show LIS metal absorption lines.

\begin{figure*}
  \includegraphics[width=\linewidth, keepaspectratio]{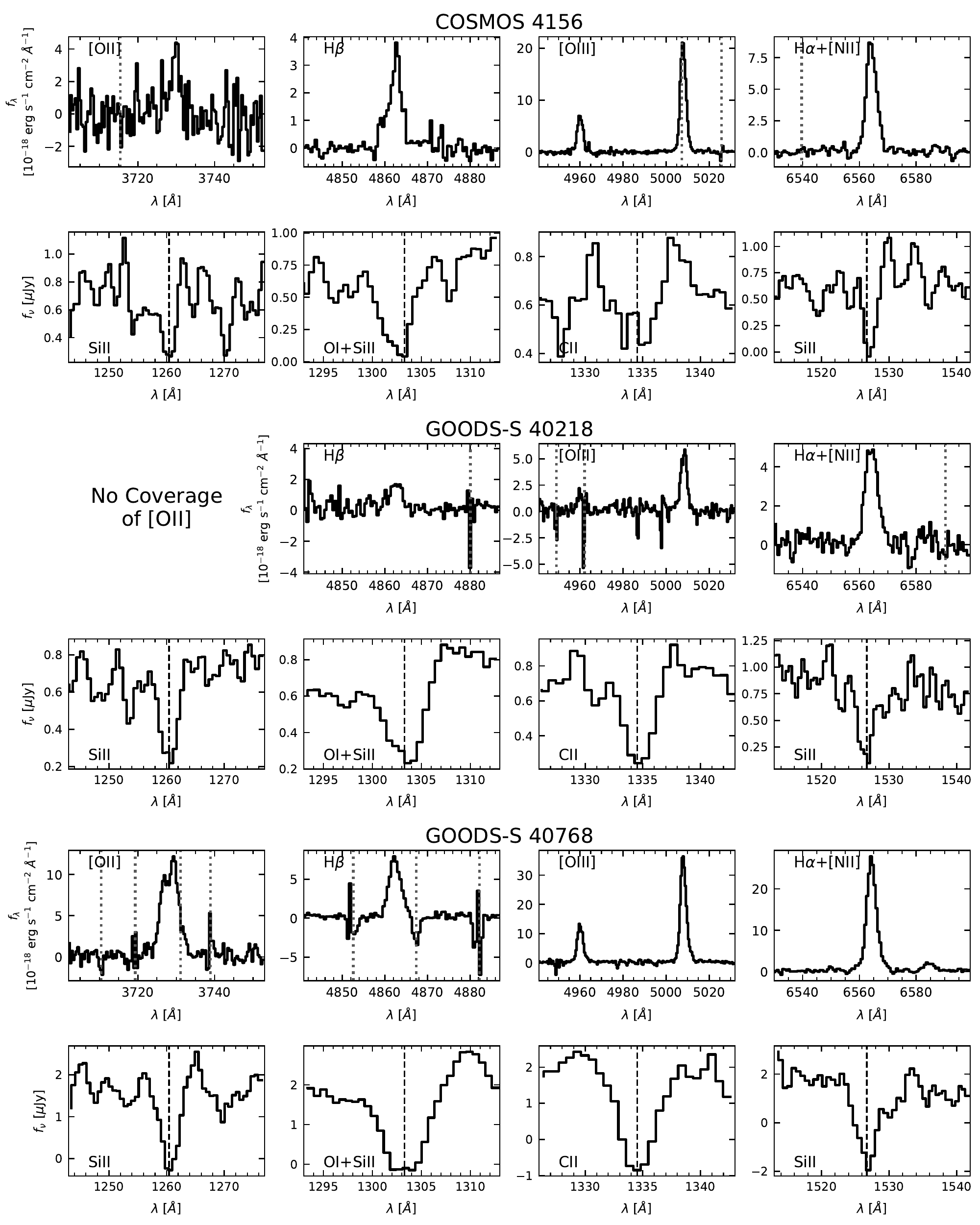}
  \vspace{-0.5cm}
  \caption{
  Plots of strong rest-optical emission and rest-UV absorption lines of the LRIS-Inflow galaxies from MOSFIRE and LRIS, respectively. 
  Dotted vertical lines mark sky lines.
  Dashed vertical lines mark the systemic absorption line centre. GOODS-S 40218 does not coverage of [OII].} 
  \label{fig:all_spec}
\end{figure*}


\bsp	
\label{lastpage}
\end{document}